\documentclass[
    10pt,
    showpacs
]{iopart}

\usepackage{iopams}

\usepackage[usenames,dvipsnames]{color}
\usepackage{graphicx,textcase} 
\usepackage{dcolumn,overpic} 
\usepackage{bm,color}
\usepackage[T1]{fontenc}
\usepackage{ae,aecompl}
\usepackage{chngcntr,gensymb}
\usepackage[para]{threeparttable}
\usepackage{etoolbox}
\usepackage{lipsum}
\usepackage[bottom]{footmisc}
\usepackage{setspace}

\AtEndEnvironment{table*}{\vskip10pt}{}{}

\usepackage{url}
\usepackage{hyperref}
\hypersetup{colorlinks=true,breaklinks,linkcolor=blue,urlcolor=blue,citecolor=blue}

\begin{document}

\title{Nature of the magnetic moment of cobalt in ordered FeCo alloy}
\author{Arsenii Gerasimov$^1$, Lars Nordstr\"om$^2$, Sergii Khmelevskyi$^{3}$, Vladimir~V.~Mazurenko$^1$ and Yaroslav~O.~Kvashnin$^2$}

\address{$^1$Theoretical Physics and Applied Mathematics Department, Ural Federal University, Mira Str. 19, 620002 Ekaterinburg, Russia}
\address{$^2$Uppsala University, Department of Physics and Astronomy, Division of Materials Theory, Box 516, SE-751 20 Uppsala, Sweden}
\address{$^3$Center for Computational Materials Science, Institute for Applied Physics, Vienna University of Technology, Wiedner Hauptstrasse 8, A-1040, Vienna, Austria}
\ead{arseniigerasimov@gmail.com}
\vspace{10pt}
\begin{indented}
\item[] \day= \today
\end{indented}

\begin{abstract}
The magnets are typically classified into Stoner and Heisenberg type, depending on the itinerant or localized nature of the constituent magnetic moments.
In this work, we investigate theoretically the behaviour of the magnetic moments of iron and cobalt in their B2-ordered alloy.
The results based on local spin density approximation (LSDA) for the density functional theory (DFT) suggest that the Co magnetic moment strongly depends on the directions of the surrounding magnetic moments, which usually indicates the Stoner-type mechanism of magnetism.
This is consistent with the disordered local moment (DLM) picture of the paramagnetic state, where the magnetic moment of cobalt gets substantially suppressed.
We argue that this is due to the lack of strong on-site electron correlations, which we take into account by employing a combination of DFT and dynamical mean-field theory (DMFT).
Within LDA+DMFT, we find a substantial quasiparticle mass renormalization and a non Fermi-liquid behaviour of Fe-$3d$ orbitals.
The resulting spectral functions are in very good agreement with measured spin-resolved photoemission spectra.
Our results suggest that local correlations play an essential role in stabilizing a robust local moment on Co in the absence of magnetic order at high temperatures.
\end{abstract}


\submitto{\JPCM}

\maketitle

\ioptwocol

\section{Introduction}
\label{intro}

The ferromagnetic (FM) alloys formed by mixing of iron and cobalt and are situated at the maximum of the Slater-Pauling (S-P) curve \cite{kubler-book}.
The latter suggests that these alloys possess the maximal possible saturation magnetization for binary $3d$ alloys (for Fe$_{1-x}$Co$_{x}$ it is reached for x$\approx$0.28 \cite{FeCo-Ms-exp}).
This fact together with extremely high magnetic transition temperatures makes this family of metals particularly promising for technological applications.
In particular, FeCo alloys would be excellent for the data storage in spintronic devices and/or permanent magnets applications if possessed larger magnetocrystalline anisotropy \cite{Burkert,turek-mae-cpa,doi:10.1002/pssb.201147100}.
Thus, nowadays a great effort is done both on theoretical and experimental fronts to achieve this by inducing tetragonal distortions, which have been predicted to vastly improve this property \cite{feco-mae-exp, FeCo-Rh,FeCo-orbmom-exp,FeCo-epitax,FeCo-C,doi:10.1002/pssb.201147100}.
In thin films, this distortion can be achieved by means of epitaxial strain, while in the bulk interstitial doping with light elements is being proposed.

From the theory side, first-principles electronic structure calculations, based on density functional theory (DFT) were quite successful in predicting the magnetic properties of Fe-Co alloys.
In 1984 Schwarz \textit{et al.} \cite{Schwarz_1984} have done a systematic study of ordered Fe-Co alloys and demonstrated that the S-P behaviour is captured by the DFT calculations and that the spin magnetic moments are in excellent agreement with experimental data \cite{FeCo-Ms-exp, moms-neutrons}.
Later this result was also confirmed for atomically disordered FeCo alloys \cite{PhysRevB.45.12911,PhysRevB.49.3352}.

\begin{figure}[!h]
\includegraphics[width=0.7\columnwidth]{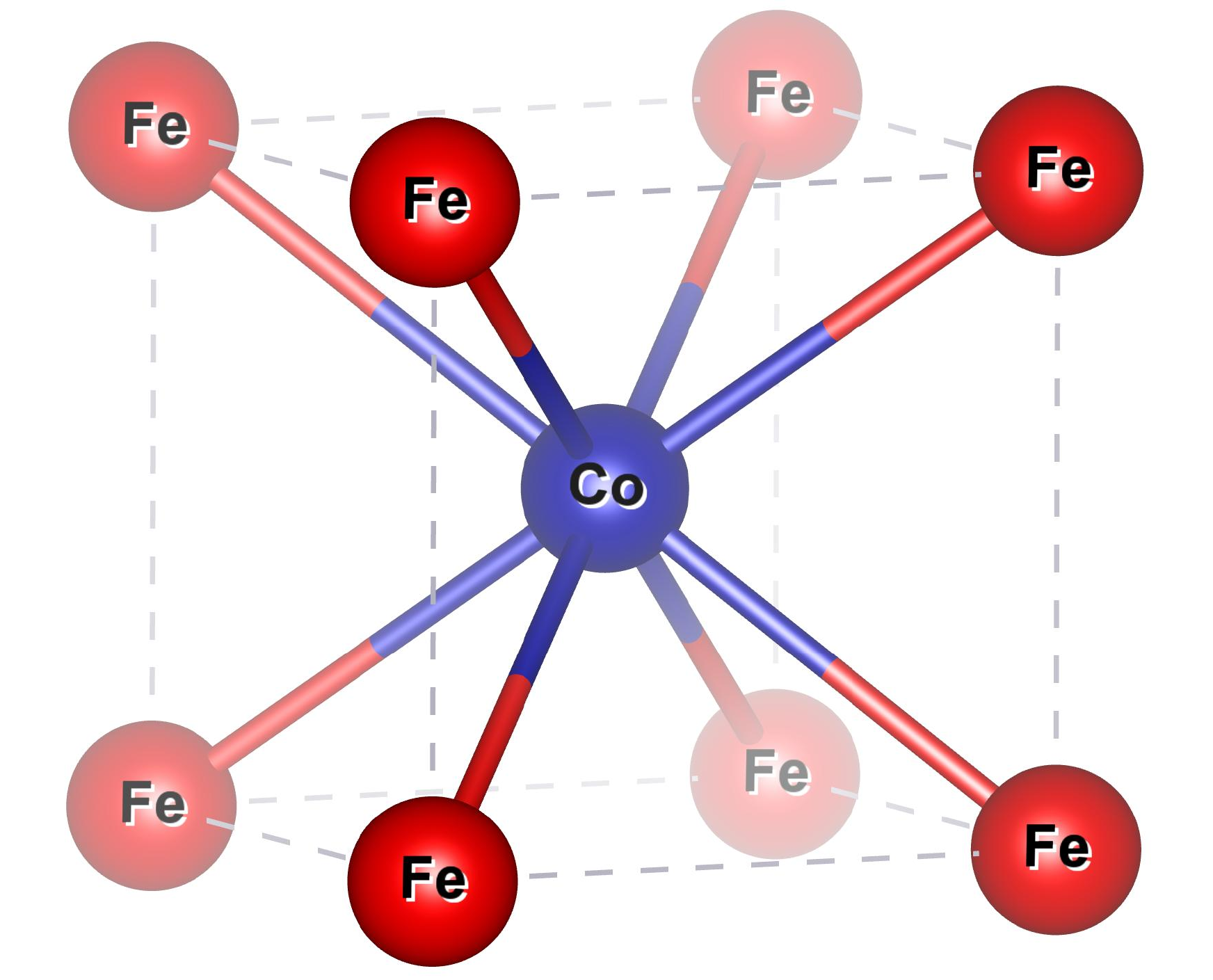}
\caption{Schematic crystal structure of the B2 FeCo alloy.}
\label{fig:str}
\end{figure}

Equiatomic FeCo alloys have a tendency to crystallize in an ordered B2 (CsCl) structure (see Figure~\ref{fig:str}) at the temperatures below 1000 K \cite{feco-phasediag, feco-phasediag2, Bozorth-book}.
In this bcc-based structure, atom A (e.g. Fe) occupies vertices of the cube and atom B (e.g. Co) sits in its centre. 
Each atom of a given type is then surrounded by eight atoms of the other type.
Upon an increase of temperature, the system undergoes an order-disorder transition to a bcc ($\alpha$) phase and remains FM.
However, at around 1230 K, another transition takes place, which brings the material into a non-magnetic fcc phase.
Hence the actual Curie temperature ($T_c$) of the material remains unknown and is often referred to as "virtual" one (see e.g. \cite{maclaren99}).
The structural stability of FeCo has been extensively studied using first-principles methods and the importance of magnetism has been reported \cite{Diaz-Ortiz06}.
Rahaman and co-workers \cite{Ruban-FeCo} have calculated the order-disorder transition temperature by extracting the effective cluster interactions from different magnetic states.
In order to model finite-temperature magnetic properties, they used partial disordered local moment (DLM) approach \cite{dlm}.
They found that the best agreement with experimental transition point is found for a reduced magnetization of $m\approx0.83$ (where $m=1$ corresponds to the FM state).
This result clearly suggests that in order to address the thermodynamic properties of this system, one needs a more rigorous theory of the thermally excited magnetic state.

The Heisenberg exchange parameters ($J_{ij}$), calculated in the FM CsCl phase, indicate a strong Fe-Co nearest neighbour interactions \cite{maclaren99,lezaic07,jakobsson13}.
The estimated $T_c$ was found to be well above 1200 K, which does not contradict experimental observations.
These estimates are based on the assumption that the exchange parameters and the magnetic moments remain relatively robust up to the magnetic transition point.
However, it has been shown that the magnetic moment of elemental bcc Fe is quite dependent on the environment \cite{chimata17}, bringing the applicability of the low-temperature $J_{ij}$'s to describe the PM state into question \cite{Attila-noncoll}.
At the same time, the magnetic excitations in elemental bcc Co show a very robust Heisenberg behaviour \cite{chimata17}.
When the two elements are mixed together, it is unclear which behaviour will prevail.

In this work, we study the nature of Fe and Co magnetic moments in the B2 FeCo alloy in an attempt to describe their behaviour at elevated temperatures. 
First, we demonstrate on the level of DFT that $m_{\mathrm{Co}}$ strongly depends on its magnetic environment, while Fe has a very stable magnetic moment.
Next, we study how these drastically different behaviours of the two constituent atoms affect the PM properties of the material.
The PM phase is simulated by employing two different models. 
The first one is the DLM approach and the second one is based on a combination of DFT and dynamical mean-field theory (DMFT) \cite{kotliar-DMFT} in order to capture the effect of strong electron correlations in the $d$ band.
We discuss the differences between the two approaches and compare the results they provide.
Finally, we investigate whether there are local moments on Co atoms in the PM phase and what is the mechanism behind its formation.

\section{Computational Details}
\label{theory}
The DFT calculations were carried out for the equilibrium experimental structure ($a$=2.856~\AA).
Most of the calculations were performed using the full-potential linearized augmented plane wave (FLAPW) method, as implemented in Elk software \cite{elk-web,singh2013}.
The spin-spirals calculations were performed by using standard DFT in the local spin density approximation (LSDA) \cite{pw92} taking advantage of the Generalized Bloch theorem \cite{Gen-Bloch-theorem}.
Those calculations were done with an increased accuracy which was achieved by setting the $R_{MT}*K_{max}$ parameter to 8.5 and $G_{max}$ in the expansion of the potential (and the density) in the interstitial to 14.0.
The non-relativistic DLM state was modelled using coherent potential approximation (CPA) \cite{cpa}, as implemented in Korringa-Kohn-Rostocker (KKR) method in the atomic sphere approximation \cite{kkr-ruban}.

The LDA+DMFT calculations were performed in several steps.
In order to take into account the correlation effects for the Fe-$d$ and Co-$d$ states, a tight-binding Hamiltonian in a localised basis was first constructed.
For this purpose, we have interfaced the Elk code with Wannier90 \cite{w90,w90-2} software, which is used to produce the maximally localized Wannier functions (MLWFs) \cite{mlwf-review}.
Once the DFT band structure was parameterized in terms of Wannier orbitals, it was used by AMULET \cite{amulet} toolbox to solve the DMFT equations.
All these steps and the relevant technical details are described in the next few subsections.

\subsection{Interface of Elk and Wannier90}
\label{elk2wannier90}
The electronic structure information provided by Elk code \cite{elk-web} is used to generate the MLWFs using Wannier90 code \cite{w90,w90-2}. 
Wannier functions are defined as the Fourier transforms of Bloch functions \cite{G.Wannier-original}:
\begin{equation}
    |W_n^{\bf R} \rangle = \frac{V}{(2\pi)^3} \int e^{-i {\bf k}{\bf R}} \sum_{m=1}^{J} U_{mn}^{({\bf k})} |\psi_{m {\bf k}}\rangle {\rm d}{\bf k}, 
\end{equation}
where $V$ is a unit cell volume, ${\bf k}$ is a wave vector inside 1st Brillouin zone, $U_{mn}^{({\bf k})}$ is a unitary matrix, $m$ and $n$ are the bands indices, $J$ is the total number of bands and $|\psi_{m{\bf k}}\rangle$ is the wave function.

There is no general rule for choosing $U_{mn}^{({\bf k})}$ and, therefore, it is necessary to impose an additional constraint on the Wannier functions, namely either to replace the action of the $U_{mn}^{({\bf k})}$ by the action of projection operator $\hat{P}_{k} = \sum_{m=1}^{J}|\psi_{m{\bf k}}\rangle \langle\psi_{m {\bf k}}|$ on the seed function $|g_{n}\rangle$ or define $U_{mn}^{({\bf k})}$ from conditions of minimizing their spread:

\begin{equation}
    \Omega = \sum_{n} [\langle W_{n}^{0}|r^2|W_{n}^{0} \rangle - |\langle W_{n}^{0}|r|W_{n}^{0}\rangle|^{2}].
\end{equation}

The latter is used for constructing MLFW \cite{Marzari_MLWF1997,Marzari_MLWFDisentangle2012}, which is currently the most widely used type of Wannier functions.

\label{elk2wannier90:Mmn}
To choose an optimal mixing matrix $U_{mn}^{({\bf k})}$ we need to calculate the gradient $dU/d\Omega$. 
The first required quantity is the overlap matrix between the periodic parts of the Bloch states $|u_{n,{\bf k}}\rangle$, calculated for the neighbouring ${\bf k}$-points:
\begin{equation}
    M_{mn}^{({\bf k},{\bf b})} = \langle u_{m,{\bf k}}|u_{n,{\bf k}+{\bf b}} \rangle,
\end{equation}
where $|u_{n,{\bf k}} \rangle = \psi_{n,{\bf k}} e^{-i {\bf k} {\bf r}}$, ${\bf b}$ is a vector that connects a given ${\bf k}$-point with its neighbours. 

The Elk code is based on the FLAPW method and the space of the unit cell is partitioned into two regions: muffin-tin (mt) spheres around each atom and the interstitial region (ir). 
Thus, the wavefunctions are explicitly separated into the two corresponding parts. 
In order to obtain the $M_{mn}^{({\bf k},{\bf b})}$ matrix, we should start with the calculation of the second-variational spinor wavefunctions in the mt and ir regions for every state of a particular ${\bf k}$-point and for every neighboring $({\bf k}+{\bf b})$-points. 
Then, on the basis of these wavefunctions, we compute the complex  charge density between a ${\bf k}$-point and its neighbor, defined as: 
\begin{eqnarray}
\rho^{({\bf k}, {\bf b})}({\bf r}) = \exp{(-i{\bf b}\cdot{\bf r})} \psi_{{\bf k}}^{*}({\bf r})\psi_{{\bf k}+{\bf b}}({\bf r}). 
\end{eqnarray}
Finally, the overlap matrix is computed as:
\begin{equation}
    \eqalign{ M_{mn}^{({\bf k}, {\bf b})} = \cr
    4 \pi Y_{0}^{0} \sum_{a} \int_{mt} {r}^2 \rho^{({\bf k}, {\bf b})}_{mt,a}({r}) d{r} + \int_{ir} \rho^{({\bf k}, {\bf b})}_{ir}({\bf r})\phi({\bf r}) d{\bf r},}
    \label{eq:Mmn_in_FLAPW}
\end{equation}
where $\rho^{({\bf k}, {\bf b})}_{mt,a}$ and $\rho^{({\bf k}, {\bf b})}_{ir}$ are the overlap charge densities inside the mt sphere around atom $a$ and in the ir region, respectively, and $\phi({\bf r})$ is a characteristic function, which cuts out the mt's (i.e. it is zero inside the muffin-tins and unity outside).

\label{elk2wannier90:Amn}
The second quantity which is needed to optimize the mixing matrix is the projection of the Bloch states $|\psi_{n,{\bf k}} \rangle$ onto some trial localized orbitals $|g_n \rangle$:

\begin{equation}
    A_{mn}^{({\bf k})} = \langle \psi_{m,{\bf k}}|g_n \rangle.
\end{equation}

The algorithm to compute $A_{mn}^{({\bf k})}$ is similar to the one described above for $M_{mn}^{({\bf k},{\bf b})}$, but instead of wavefunctions for $({\bf k} + {\bf b})$-points a set of trial wave functions for every state of a particular ${\bf k}$-point is used. 

\label{elk2wannier90:TrialWF}
We chose to define the trial wavefunctions as the first APW function from the DFT basis set. They are centered at a given atom, localized and have a pure angular momentum character. 
Linear combinations including various hybrid orbitals such as $sp^3$, which are supported by the \textit{Wannier90 library}, can also be formed.

We should note that the MLWF can be extracted from other FPLAPW-based codes and the present implementation is not unique in this sense.
For instance, Wannier90 has been interfaced with FLEUR \cite{Freimuth_fleur2wann} and Wien2k \cite{KUNES_wien2wann} codes.
Our implementation is similar to those and is even more straightforward, since it is based on the direct calculation of the generalized density $\rho^{({\bf k}, {\bf b})}$.

\subsection{Wannierization}
\label{wannierization}

In order to build an effective Hamiltonian in the MLWF basis, we have used the band structure obtained from non-magnetic LDA calculations using (25,25,25) Monkhorst-Pack ${\bf k}$-point grid. 
Once the potential is converged and the positions of the bands are accurately identified, the wannierization can be efficiently performed using a coarser ${\bf k}$-mesh.
Here, the Wannier functions were built using the Kohn-Sham states on a (16,16,16) ${\bf k}$-point mesh. 
We projected the bands on the orbitals of $s$, $p$ and $d$ characters of iron and cobalt. 
Thus, we had 18 projections in total, since the states are spin-degenerate. 

Since the bands near the Fermi level are not isolated, \textit{i.e.} we are dealing with partially occupied bands of a metal, a disentangling procedure \cite{wannier_disentangl} has to be applied. 
We took a substantial amount of unoccupied bands (42 in total) and specified the energy windows for the disentanglement procedure. 
The first (outer) window covers all the 42 bands. 
The second window, the so-called frozen "inner" window, spans from -8.5 eV up to 1 eV (relative to Fermi level) and contains Fe and Co \textit{3d} states.

\subsection{DMFT calculation} \label{dmft_setup}
The wannierized band structure of non-spin-polarized calculation was used for setting up the non-interacting Green's function for the impurity problem.
The magnetic DMFT calculations were done in the LDA+DMFT manner, where the magnetic field is used to break the spin degeneracy of the starting self-energy $\Sigma$. Thus the magnetism comes entirely from $\Sigma$, while the non-interacting Hamiltonian remains non-magnetic.
This recipe is adopted from ~\cite{DC_sigma_zero}.

The DMFT calculations were carried out to capture strong correlations between the \textit{d} electrons of iron and cobalt atoms. We used the following Hubbard parameters $U_{\mathrm{Fe}}$ = 2.6 eV and $U_{\mathrm{Co}}$ = 3.7 eV. The Hund's exchange parameter was set to $J$ = 0.9 eV for both atomic species. 
These values were extracted from X-ray photoelectron spectra \cite{FeCo.experimentUandJ}.

We used the segment version of continuous-time hybridization expansion quantum Monte Carlo (CT-QMC-HYB) solver \cite{Gull_CT-QMC-HYB} to treat the impurity problem. In the case of FeCo, the impurity problem has to be solved twice at each iteration: one impurity problem for iron and one for cobalt. 

Since part of Coulomb interaction is already taken into account in the LDA Hamiltonian, some interaction contributions would be counted twice within the LDA+DMFT scheme. 
So we have to avoid a so-called double-counting (DC) problem. 
Here we have used a DC correction, which is based on Friedel sum rule and postulates that the total number of correlated electrons on each impurity site is the same as in the non-interacting problem. 
In practice, we chose the DC chemical potential in such a way that the total impurity occupation coincided with that in LDA calculation \cite{PhysRevB.77.205112}.

In our calculations, the DC corrections for iron and cobalt were set to $\mu_{dc}(\mathrm{Fe})=$-12.2 eV and $\mu_{dc}(\mathrm{Co})=$-21.2 eV, respectively.
We also tried FLL type of DC correction instead of fixing it on the certain values, but it led to a strong deviation of the occupation and thus to unphysical values of the fluctuating moments. 
To perform analytic continuation of the Green's and spectral functions from the Matsubara to real energy axis, we used the maximum entropy method (MEM) \cite{Sandvik_maxent, Gubernatis_maxent, Silver_maxent}. 

\section{Results and Discussion}
\label{result}

\subsection{LSDA}
\label{LSDA_results}
\begin{figure}[!h]
\includegraphics[width=\columnwidth]{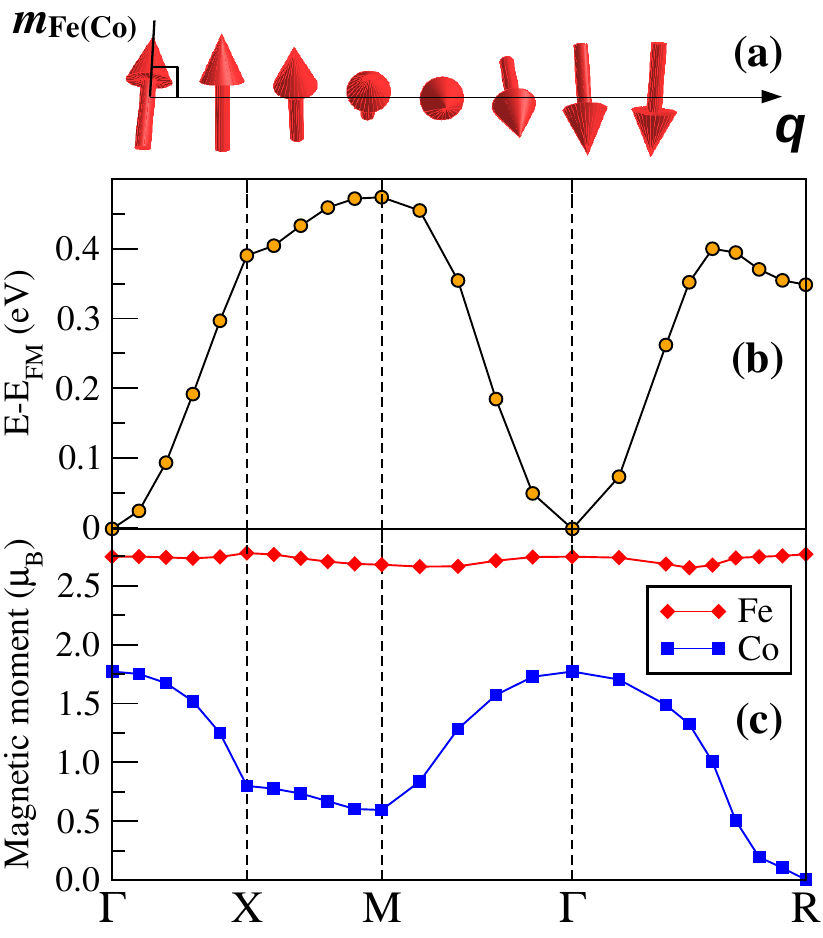}
\caption{(a) Considered family of spin spiral states with the propagation vector being perpendicular to the magnetization. (b) Relative total energies of the considered spin spiral states with respect to that of the FM state. (c) The magnitude of Fe and Co magnetic moments in the corresponding configurations.}
\label{fig:spirals}
\end{figure}

We began our study by addressing the question of how localized (or itinerant) the magnetic moments in the ordered FeCo alloy.
This can be verified by performing the calculations for various hypothetical magnetic orders. 
From \textit{ab initio} calculations perspective, if a moment is strongly dependent on the orientation of its neighbouring spins, it is a signature of that it has itinerant character.
In order to effectively explore a large number of different magnetic configurations, we employed spin spiral calculations.
Here we considered planar spin spirals, which means that the Fe and Co magnetic moments rotate in the plane perpendicular to the propagation vector of the spin spiral ${\bf q}$.
The initial phase shift between Fe and Co moments is commensurate with the chosen ${\bf q}$ and thus it can be viewed as a single spin-spiral propagating through both sublattices (see Figure~\ref{fig:spirals}(a)).

The results of these calculations are shown in Figure~\ref{fig:spirals}.
In the panel (b), the total energy as a function of the spin spiral wavevector ${\bf q}$ is shown. 
The minimum at $\Gamma$ point and a parabolic dependence in its vicinity clearly indicates a strong tendency to ferromagnetism.
There is a local minimum at the $R$ point ($ {\bf q}=(1/2,1/2,1/2)$), which corresponds to the AFM(111) solution. 
In order to understand this result, we should analyze the data shown in the panel (c) of the same figure.
Here we show the self-consistent values of the Fe and Co magnetic moments ($m_{\mathrm{Fe}}$ and $m_{\mathrm{Co}}$, respectively), obtained in these calculations.
First of all, one can see that $m_{\mathrm{Fe}}$ is practically independent of ${\bf q}$ and roughly retains its value corresponding to the FM state (2.75 $\mu_B$).
At the same time, $m_{\mathrm{Co}}$ shows a strong variation. In particular, for the R point the Co moment completely vanishes.

It is a common situation when the "stiffness" of an atomic moment is defined by whether its majority band is partially or completely filled.
In the former case corresponds to the so-called "weak" ferromagnets (see e.g. \cite{P.Mohn-magnetism-book}), which emphasizes the fact that the moment is more susceptible to various external stimuli, like stress, magnetic field, etc.
However, we note that this argument can not explain the different behaviours of $m_{\mathrm{Fe}}$ and $m_{\mathrm{Co}}$, we find here.
In fact, partial DOS calculated for the FM state shows that both Fe and Co majority states are occupied \cite{maclaren99}.

Here we see that the magnetic moment of Co shows itinerant behaviour. 
For instance, the magnetic moments of Pt or Pd atoms in Co-Pt and Fe-Pd alloys show similar sensitivity to the magnetic environment (see e.g. \cite{polesya}).
The latter is clear from the fact that these elements are non-magnetic in their pure elemental forms, but acquire induced magnetization when placed next to $3d$ element. 
Elemental Co, on the other hand, is a well-behaving ferromagnet and therefore we believe the interpretation is slightly different here.

It has been shown previously \cite{maclaren99,lezaic07,jakobsson13} that the nearest-neighbour Fe-Co exchange interaction ($J_\mathrm{Fe-Co}$) in the ordered FeCo alloys is very strong and is FM.
When the ${\bf q}$ vector of the spin spiral is set, it defines a certain angle between neighbouring $m_{\mathrm{Fe}}$ and $m_{\mathrm{Co}}$.
For an angle $\Theta$, the energy cost of such spin spiral configuration with respect to the FM solution becomes: $zJ_\mathrm{Fe-Co}m_{\mathrm{Co}}m_{\mathrm{Fe}}(1-\cos(\Theta))$, where $z$ is the number of the nearest neighbours.
If this energy penalty is larger than the energy gained due to the formation of the moment, it is energetically favourable to decrease or even to shrink one of the moments.
Since the exchange splitting of Fe is larger than that of Co \cite{Schwarz_1984}, the latter can be sacrificed at a smaller cost. 

The emergence of the local minimum in Figure~\ref{fig:spirals}(b) at the $R$ point is related to this trade-off between inter-atomic and intra-atomic exchange, described above.
If the Co moment was still finite, the relative energy of this state would be much higher. 
The same argument applies to all other spin spiral states, where the $m_{\mathrm{Co}}$ is substantially reduced as compared to its value in the FM state.
One can notice, however, that $m_{\mathrm{Co}}$ does not go to zero for the $X$ and $M$ high-symmetry points, as it does for $R$ point.
In all these three states, the angle between nearest Fe and Co moments is 90\degree, but the difference is in the relative orientation of the Co moments on different sites.
For the spin spirals with $R$ wavevector, $m_{\mathrm{Co}}$ is antiparallel with respect to all six next-nearest neighbour Co moments.
Since the corresponding coupling $J_\mathrm{Co-Co}$ is weak, but also ferromagnetic \cite{maclaren99,jakobsson13}, the formation of the moment is completely unfavorable.
On the contrary, for the \textbf{q} corresponding to $X$ and $M$ points, $m_{\mathrm{Co}}$ is ferromagnetically aligned at least to some of its second neighbours spins~\footnote{For the $X$ ($M$) point, four(two) out of six next-nearest neighbour spins have a parallel orientation with respect to that of the selected atom.}.
Therefore, the $J_\mathrm{Co-Co}$ comes into play and competes against the $J_\mathrm{Fe-Co}$-derived term.
This results in finite Co magnetization for these spin spiral states.

Thus, it is shown that the strong dependence of the Co moment on the spiral wavevector is actually related to the exceptionally strong exchange coupling between Fe and Co (relative to the on-site intra-atomic exchange on Co).

\subsection{Disordered local moments}
Next, we investigated whether such sensitivity of Co moment to its magnetic environment has implications for the magnetically disordered phase.
A widely used model of the PM state is based on the disordered local moment approximation.
It can be realized by considering the following binary alloy Fe$^{\uparrow}_{0.5}$Fe$^{\downarrow}_{0.5}$Co$^{\uparrow}_{0.5}$Co$^{\downarrow}_{0.5}$, where each sublattice is occupied with equal probability by Fe or Co atoms with the magnetic moments pointing along/opposite to the direction of the field.
The averaging is done by using CPA theory, which is the best single-site approach for modelling uncorrelated disorder.

The calculated DLM magnetic moments are equal to 2.6 $\mu_{B}$ for iron, which agrees with LSDA results for ferromagnetic configuration (2.66 $\mu_B$ obtained with the KKR method).
At the same time, the magnetic moment on cobalt shrinks to zero, which is usually a sign of the itinerant character of magnetism.
Such a behaviour has been reported before for B2 FeCo, but not analyzed in detail \cite{comtesse-thesis}.

\begin{figure}[!h]
\includegraphics[width=\columnwidth]{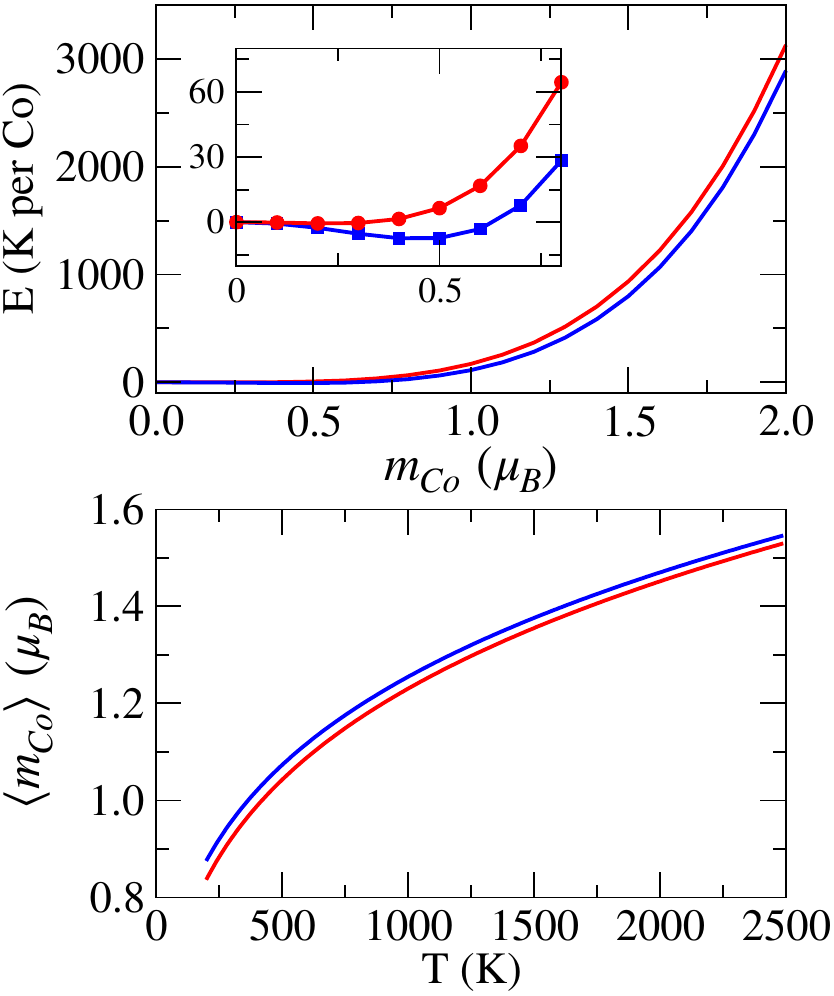}
\caption{Top panel: Relative total energies (in K per Co atom) of the DLM calculations with various prescribed values of the Co magnetization (in $\mu_B$ per atom). Bottom panel: DLM-derived temperature-dependent Co local moment, obtained using ~\ref{e-lsf}. Red colour - LSDA, Blue colour - GGA results.}
\label{fig:dlm}
\end{figure}

In order to get insight into this behaviour, we have performed a series of DLM calculations with the constrained magnetic moment on Co site.
The results are shown in Figure~\ref{fig:dlm}. 
The calculations were performed using either LSDA or generalized gradient approximation (GGA) \cite{gga-pbe} for the exchange-correlation energy.
For LSDA-derived results, the total energy has a minimum at zero Co moment, which corresponds to the global minimum of the unconstrained DLM calculation.
Thus we see a manifestation of the same behaviour of $m_{\mathrm{Co}}$ that we found for the magnetically ordered states presented in subsection \ref{LSDA_results}.
When Co magnetic moment has a non-magnetic surrounding, it tends to vanish.
However, the situation is slightly different for the GGA functional. The total energy profile is extremely shallow, but has a minimum at a finite value of the Co magnetization, roughly 0.45 $\mu_B$. This value is still substantially smaller than that in the FM ground state ($\sim$1.75 $\mu_B$).

Within the DLM framework, a finite magnetic moment on Co in the PM state emerges from the temperature-induced longitudinal spin fluctuations \cite{moriya-book}. This effect can be captured using a fixed spin moment formalism on top of the DLM state \cite{PhysRevB.75.054402}.
For instance, taking into account longitudinal fluctuations in addition to the Heisenberg interactions has been shown to substantially affect the estimates of the $T_c$ of disordered Fe$_{1-x}$Co$_x$ alloys \cite{PhysRevB.95.184432}.
Here we apply this formalism to improve the description of magnetism of Co.

The converged constrained solutions, shown in the top panel of Figure~\ref{fig:dlm} can be interpreted as excited states of paramagnetic FeCo. 
These states will become accessible at finite temperature and will induce finite magnetization on the Co site.
We calculated the temperature-dependent local moment of Co atom using the scheme described in details in ~\cite{PhysRevB.94.024420}. 
Using the DLM total energies shown in the top panel of the Figure~\ref{fig:dlm}, the thermal average of the local moment amplitude on the Co sites can be calculated as follows:
 \begin{eqnarray}
 \langle m(T) \rangle = \frac{\int m^3 \exp{(-E_{DLM}(m)/k_BT)} dm}{\int m^2 \exp{(-E_{DLM}(m)/k_BT)} dm}
 \label{e-lsf}
 \end{eqnarray}
According to the obtained results, shown in Figure~\ref{fig:dlm} (bottom panel), the magnetic moment on cobalt shows a pronounced temperature dependence. 
It shows no signs of approaching the saturation even at higher temperatures (above 2000K).

Finally, we compared the total energies of the DLM and FM states, calculated using the KKR method. We found that the DLM solution is 0.46 eV larger than the FM one, which corresponds to an effective temperature of $\approx$ 5300 K. This clearly indicates that the DLM approach does not provide an adequate description of the paramagnetic state of FeCo. 

There might be several reasons for such behaviour, such as the assumption of uncorrelated neighbours within DLM calculation or incomplete description of the correlation effects.
The former reason is indeed plausible, but then would imply that Co moment is finite above $T_c$ only thanks to short-range order.
However, additional calculations performed with DLM$+U$ clearly indicates the importance of local correlations.
For these calculations, we adopted the same choice of $U$ and $J$ as specified in Section IIC.
As a result, we obtained local magnetic moments of 2.95 and 1.63 $\mu_B$ per Fe and Co atom, respectively. Thus, taking into account Hubbard $U$ drastically changes the behaviour of Co moment.

In the next section, we will provide a more complete description of the material in the presence of strong dynamical correlations.

\subsection{LDA+DMFT}
The relation between DLM and LDA+DMFT approaches has been discussed in ~\cite{PhysRevB.67.235105}. 
Even though both methods provide dynamical self-energies, the underlying physical mechanisms are different.
DLM captures spin fluctuations in a detailed way, but the quantum nature of spin and Hubbard correlations are completely neglected. As a result, the conditions for the formation of the local moment are different in the two methods. 
In the DLM approach, which is based on LSDA, the magnetism is governed by the effective Stoner $I$ entering the functional, whereas in LDA+$U$/DMFT this part is played by Hubbard $U$ \cite{Anisimov91}.

\begin{figure}[!h]
\includegraphics[width=1.00\columnwidth]{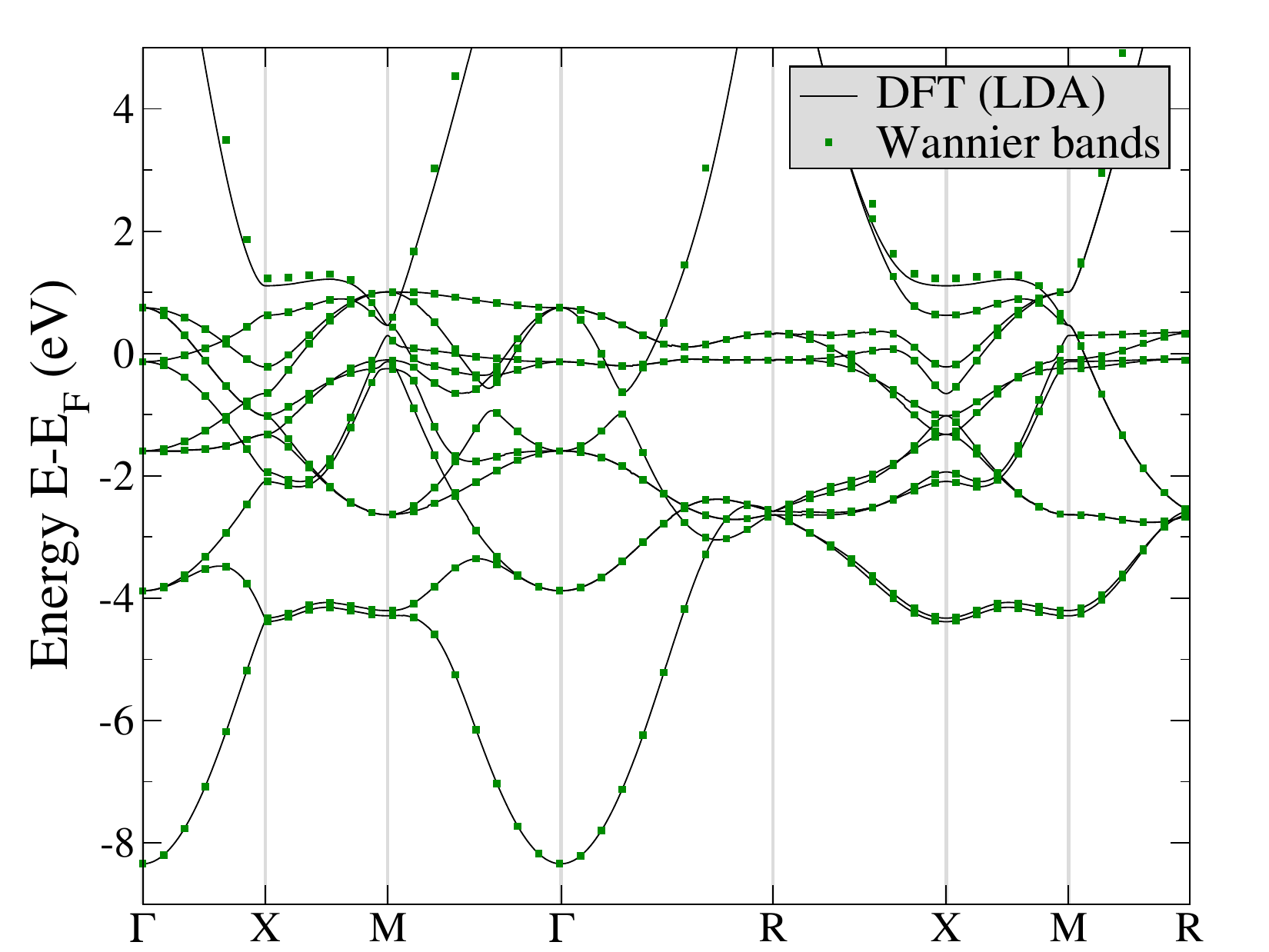}
\caption{Comparison of non-magnetic band structures of B2 FeCo alloy calculated using the full-potential linearised augmented-plane-waves (black lines) and low-energy model in Wannier functions basis of \textit{spd} character (green dotted lines). The Fermi level corresponds to the zero energy.}
\label{fig:FeCo.LDAvsWannier.Bands}
\end{figure}

In order to examine the impact of dynamical correlation effects on the paramagnetic moments, we performed LDA+DMFT calculations.
Figure~\ref{fig:FeCo.LDAvsWannier.Bands} shows the LDA-derived band structure of non-magnetic B2 FeCo alloy. 
To parametrize the LDA Hamiltonian, we have constructed the low-energy model in the Wannier functions basis. 
MLWFs of \textit{spd} character were computed as described in \ref{elk2wannier90}. 
From Figure~\ref{fig:FeCo.LDAvsWannier.Bands} one can see that the wannierized band structure reproduces that of LDA extremely well. 
Using the constructed low-energy model we have performed paramagnetic and ferromagnetic LDA+DMFT calculations for FeCo.

\begin{figure*}[!t]
\includegraphics[width=1.8\columnwidth]{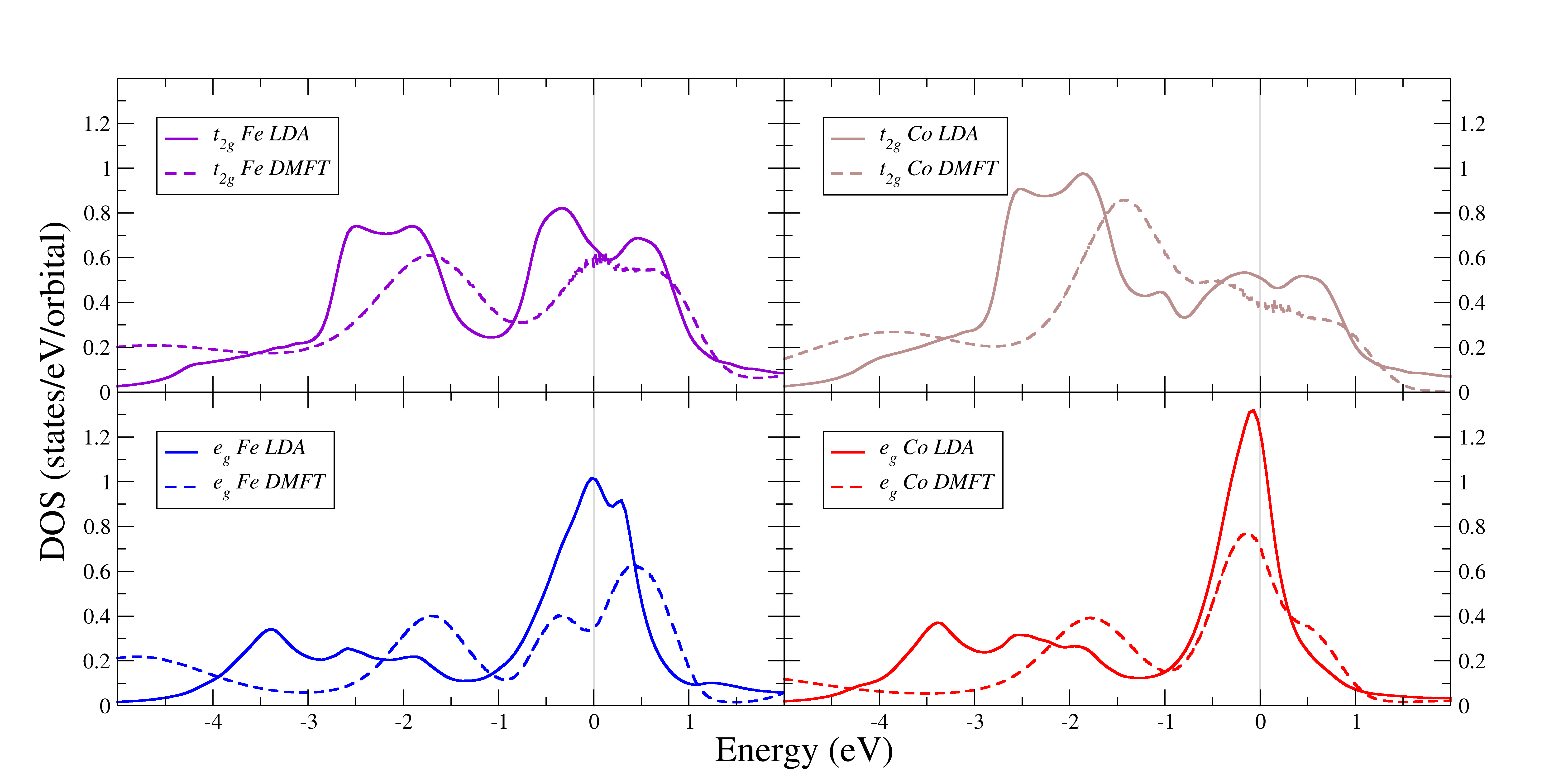}
\caption{Partial $t_{2g}$ (upper half) and $e_{g}$ (bottom half) densities of states of paramagnetic B2 FeCo alloy obtained by LDA and LDA+DMFT. 
Both sets of results are broadened with the same thermal smearing of 1450 K. The LDA+DMFT calculations were performed with $U_{\mathrm{Fe}}$ = 2.6 eV, $U_{\mathrm{Co}}$ = 3.7 eV and $J$ = 0.9 eV. The Fermi level (vertical line) corresponds to the zero energy.}
\label{fig:FeCo.LDAvsMaxent}
\end{figure*}

In Figure~\ref{fig:FeCo.LDAvsMaxent} we present paramagnetic LDA+DMFT results in comparison with LDA ones. The DMFT simulations were carried out at $T$=1450 K. 
One can see that the LDA+DMFT spectral functions obtained at finite high temperatures are much more smeared in comparison with LDA results.
As typical for transition metals, the dynamical Coulomb correlations in FeCo lead to the renormalization of electronic spectra near the Fermi level \cite{anisimov2010electronic}, which changes the effective mass of the charge carriers. This is a manifestation of the effect of Hubbard $U$, which can be observed experimentally, for example, in (angular-resolved) photoemission spectroscopy. 
We will later present a comparison of our results with the measured data for ferromagnetically ordered phase.

\begin{figure*}[!t]
\includegraphics[width=1.8\columnwidth]{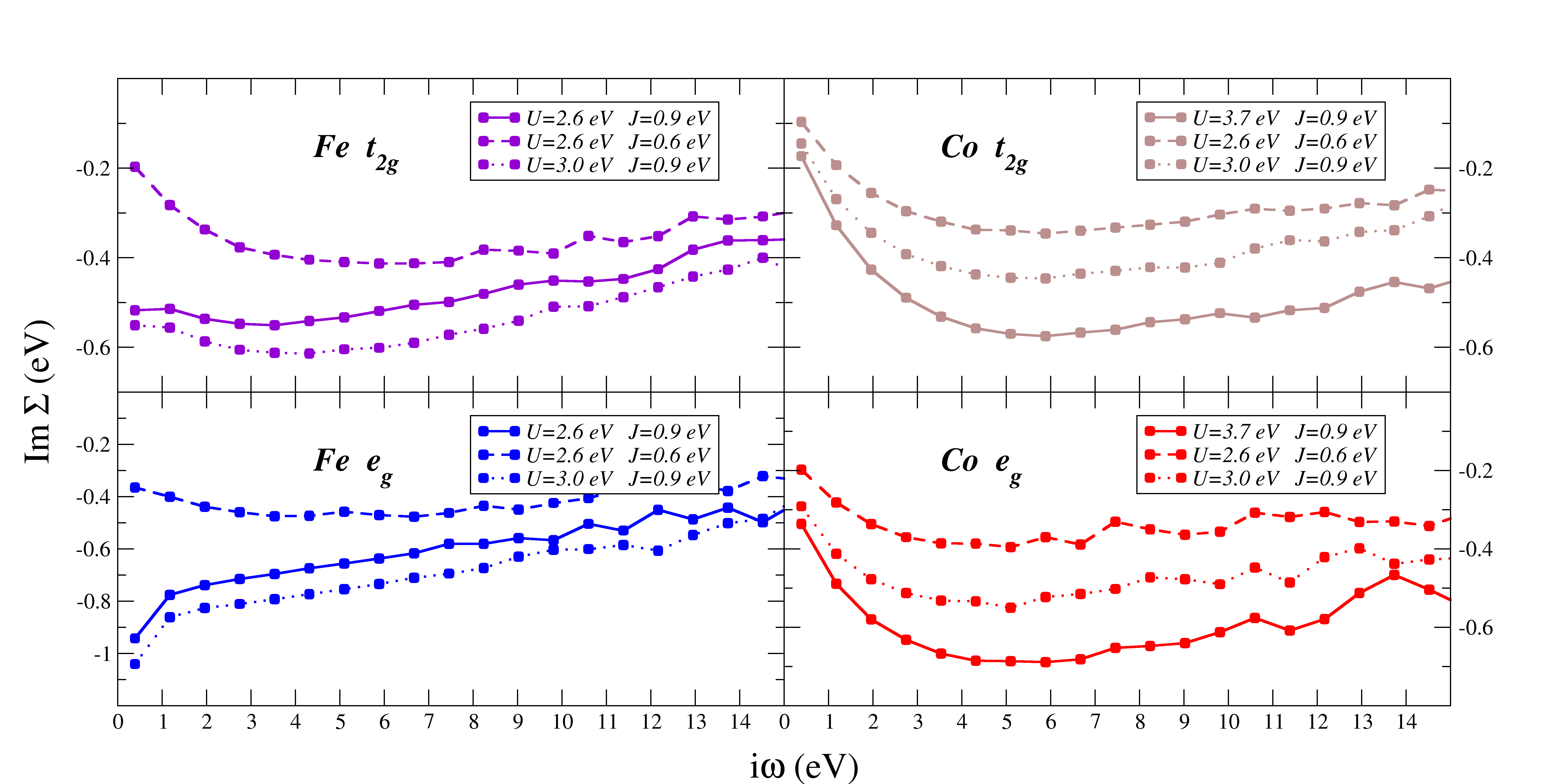}
\caption{Imaginary part of the self-energy of $t_{2g}$ (upper half) and $e_{g}$ (bottom half) states plotted on imaginary (Matsubara) axis obtained from LDA+DMFT with different $U$ and $J$ parameters.}
\label{fig:FeCo.Imag_self_enrgy}
\end{figure*}

From the imaginary part of the self-energies, presented in Figure~\ref{fig:FeCo.Imag_self_enrgy}, a divergent-like behaviour was discovered for iron $t_{2g}$ and $e_g$ states, the latter being similar to the case of pure bcc Fe \cite{Belozerov2014,Katanin2010bccFe}. 
This means that there is a very strong damping of Fe-derived quasiparticles.
At the same time, the 
$e_g$ and $t_{2g}$ of Co show Fermi-liquid character.
We have calculated the quasiparticle weights of these states using the following expression \cite{PhysRevB.75.045125}:
 \begin{eqnarray}
 Z = \biggl(1-\frac{\mathrm{Im} \Sigma(i\omega_0)}{\omega_0}\biggl)^{-1},
 \label{q-p-weight}
 \end{eqnarray}
where $\omega_0$ is the first Matsubara point.
The following values have been obtained:
$Z_{\mathrm{Co}-{e_g}}$ = 0.54 and $Z_{\mathrm{Co}-{t_{2g}}}$ = 0.70.
Thus, even though the quasiparticles originating from Co orbitals are well defined, their masses are substantially renormalized due to dynamical correlations.

In order to check the influence of $U$ and $J$ parameters on these results, we performed two series of additional calculations with the following parameterization: $U_{\mathrm{Fe,Co}}$ = 3 eV and $J_{\mathrm{Fe,Co}}$ = 0.9 eV; $U_{\mathrm{Fe,Co}}$ = 2.6 eV and $J_{\mathrm{Fe,Co}}$ = 0.6 eV. 
It is clearly seen in Figure~\ref{fig:FeCo.Imag_self_enrgy} that by decreasing $J$ the Fermi-liquid character of iron $t_{2g}$ and $e_g$ states restores (the last also agrees with elemental iron studies \cite{Belozerov2014,Katanin2010bccFe}). 
Meanwhile, the variation of $U$ does not affect the results dramatically.

The next step of our investigation was to simulate the ferromagnetically ordered FeCo alloy with LDA+DMFT. This was done by applying an external magnetic field of 0.01 eV on the level of the Hubbard model. For T=1000 K (the lowest temperature in our LDA+DMFT calculations), it was found that the following magnetic moments are stabilized: 2.69 $\mu_{B}$ (iron atoms) and 1.82 $\mu_{B}$ (cobalt atoms).
These values are in good agreement with previous theoretical and experimental studies \cite{Victora1984_calcMagnetizationFeCo,Bardos_MeanMomentsFeCo,Burkert_GiantAnisotropyFeCo} as well as with the results of LSDA calculations, presented in subsection~\ref{LSDA_results}.
We additionally performed simulations applying local magnetic fields selectively on either iron or cobalt atoms. These calculations also resulted in the finite magnetization of Co site, asserting that this moment is intrinsically stabilized.

\begin{figure*}[!t]
    \begin{minipage}[b]{0.49\linewidth}
        \includegraphics[width=1.\columnwidth]{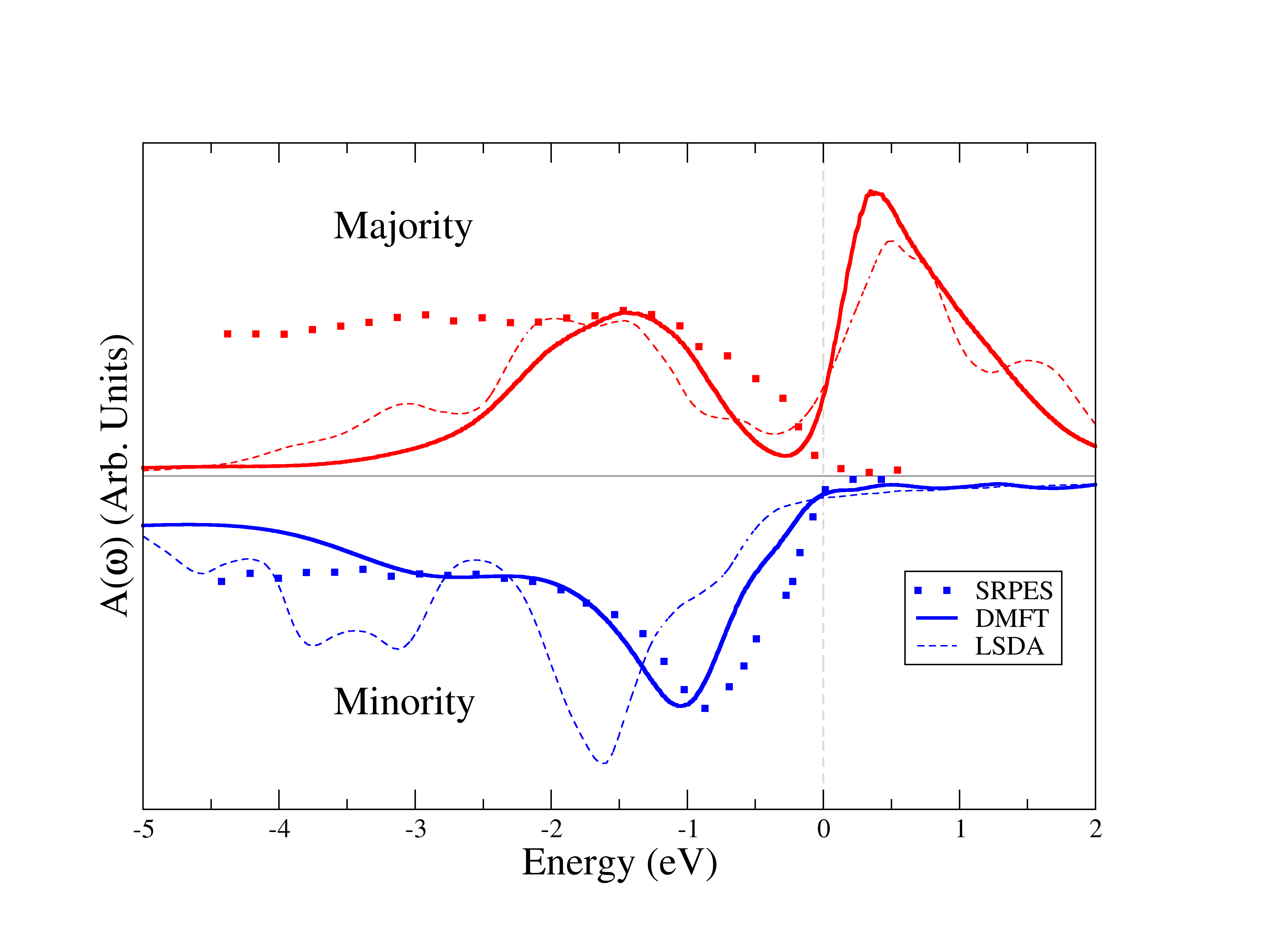}
        (a)
    \end{minipage}%
    \begin{minipage}[b]{0.49\linewidth}
        \includegraphics[width=1.\columnwidth]{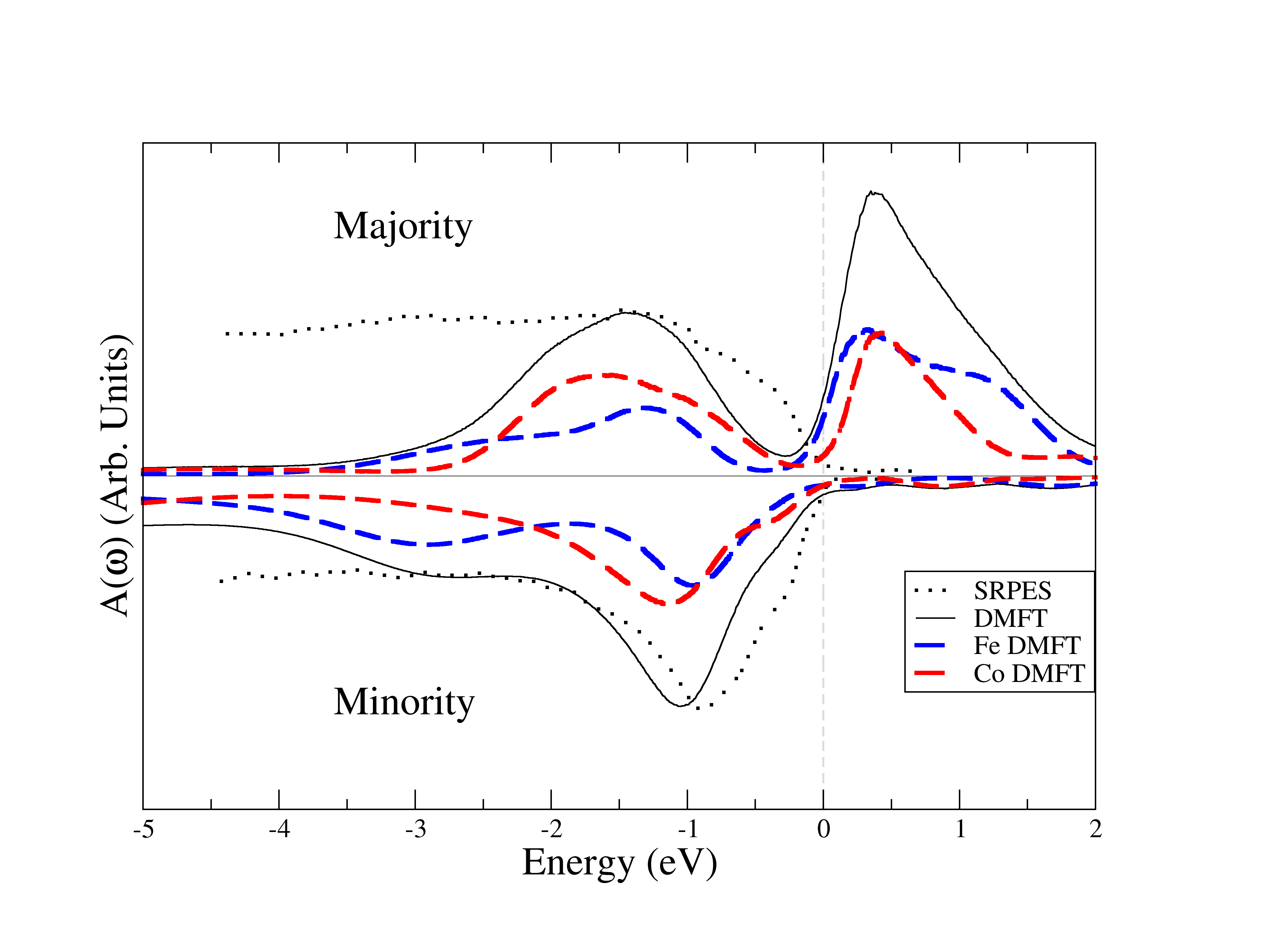}
        (b)
    \end{minipage}%
    \caption{(a) Spin-resolved density of states of ferromagnetic B2 FeCo alloy obtained by LSDA (dashed lines), LDA+DMFT (solid lines) and SRPES (dotted lines). (b) Comparison of SRPES and LDA+DMFT spectra with different contributions of $d$-states of Fe and Co atoms obtained in our calculations.}
    \label{fig:FeCo.DMFTvsSRPES}
\end{figure*}

Figure \ref{fig:FeCo.DMFTvsSRPES}(a) shows spin-resolved photoemission spectroscopy (SRPES) data~\cite{SRPES} along with the DMFT DOS obtained in this work. Experimental measurements were performed in bcc Fe$_{0.5}$Co$_{0.5}$/MgO thick films at room temperature using photon energy $h\nu$ = 40 eV and $s$-polarizations of the incoming photons. The DMFT results were obtained at $T$=1450 K with the standard for this work choice of $U$ and $J$ parameters (see \ref{dmft_setup}). 
The inspection of the figure reveals that LSDA calculations fails in describing the experimental spectra. The most indicative situation is with minority states. The LSDA-derived peak located at -1.65 eV does not match the position of the experimental one, but their widths are similar. 
At the same time, the peaks in the interval spanning from -4 eV up to -3 eV, predicted by LSDA, are not visible in SRPES results.
On the other hand, main features of experimental spectra are well reproduced in our LDA+DMFT calculations. From Figure \ref{fig:FeCo.DMFTvsSRPES}(b) we can observe a slightly enhanced agreement of experimental data and our intensity at the peak near -1.35 eV of spin-up states, which is formed with slight dominance of $d$-orbitals of Co atom, and much better agreement of spin-down states peak centered at -0.85 eV, which is formed with almost equal contributions of $d$-states of both elements.
We believe that such a dramatic improvement in the agreement with experimental spectra clearly suggests that taking into account dynamical correlations is inevitable for describing the electronic structure of FeCo alloy.

\begin{figure}[!h]
\includegraphics[width=\columnwidth]{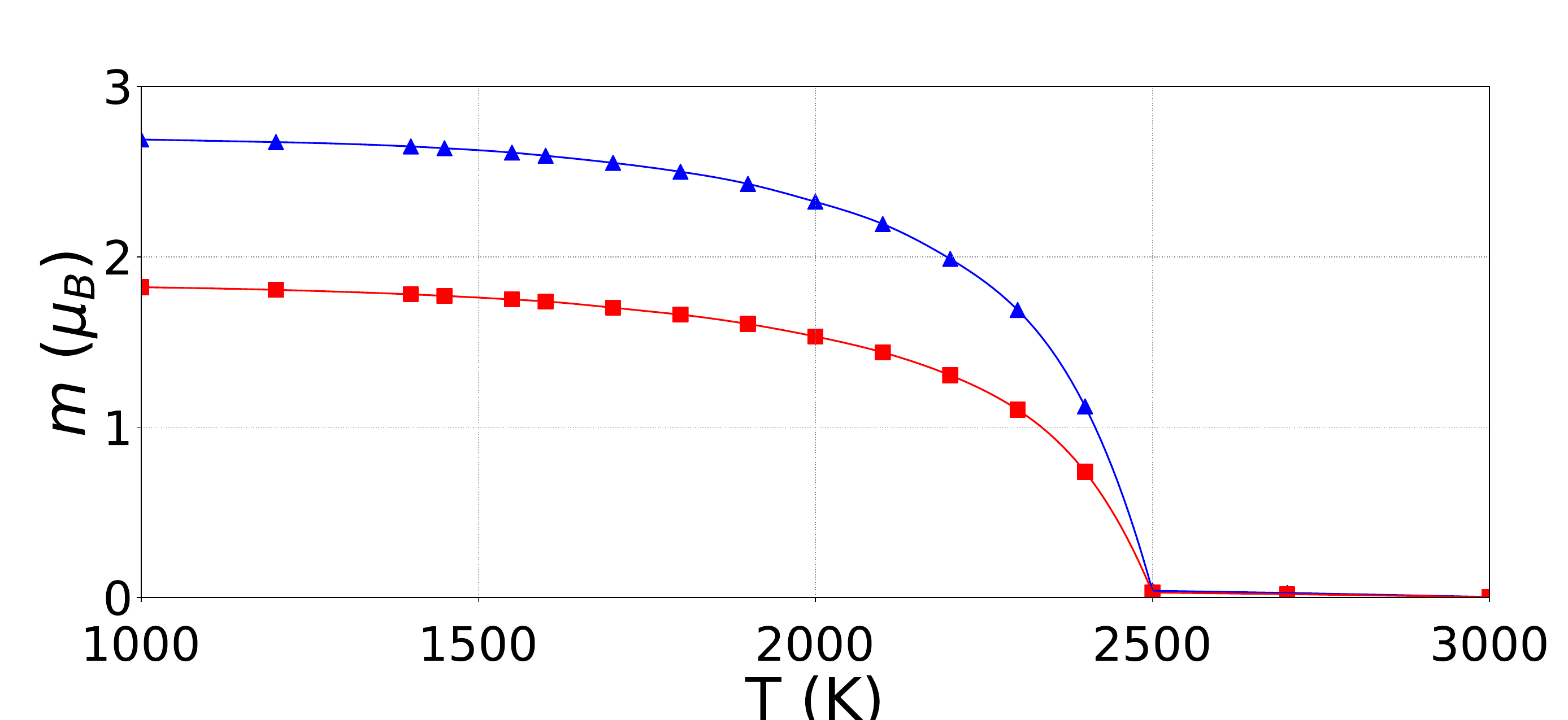}
\caption{LDA+DMFT-derived ordered magnetic moments on Fe (blue line) and Co (red line) atoms in the B2 phase of FeCo alloy as a function of temperature.}
\label{fig:FeCo.DMFTmagMom}
\end{figure}

In contrast to LSDA and DLM,  LDA+DMFT can be used to explore the \textit{explicit} thermal properties of a material in question. For instance, one can detect its Curie temperature. This was done using the following procedure. First, we stabilized the magnetic solutions for the B2 FeCo system by applying an external magnetic field of 0.01 eV at different temperatures from 1000 K to 4000 K. Then, starting with the obtained spin-polarized self-energies we continued the calculations without a magnetic field. It was found that below 2499 K each solution remained spin-polarized. At the same time, at T > 2499 K the magnetizations of both atoms shrink to zero (see Figure~\ref{fig:FeCo.DMFTmagMom}). Thus, this temperature can be associated with the Curie temperature. 

The overestimation of the experimental value of the $T_c$ by 60\% can be explained by the fact that we use the density-density form of the Coulomb interaction $U$ in the DMFT calculations that breaks the rotational invariance of the system. 
The latter has been shown to result in an overestimation of the magnetic transition temperatures in the Hubbard model using DMFT in particular \cite{PhysRevB.74.155102,Antipov2011,PhysRevB.86.035152,Belozerov2014}.

\begin{figure}[!h]
\includegraphics[width=\columnwidth]{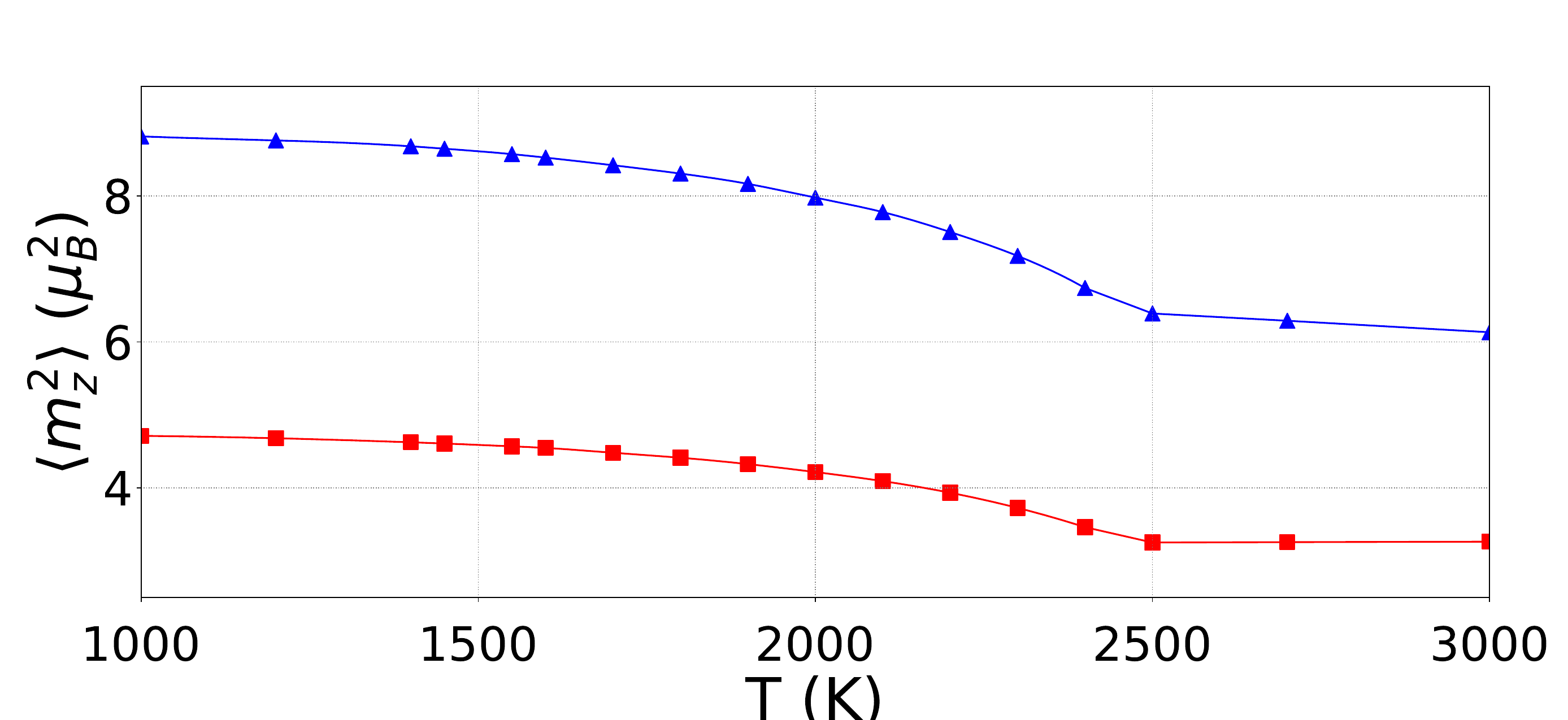}
\caption{Instant squared $z$-projected magnetic moments on Fe (blue line) and Co (red line) atoms in B2 FeCo alloy as a function of temperature from LDA+DMFT calculation.}
\label{fig:FeCo.DMFTmagMomSquared}
\end{figure}

In addition to the projected magnetization, a many-body approach like LDA+DMFT allows us to extract the values of fluctuating magnetic moments. 
For this purpose, we have calculated the average value of the squared magnetic moment operator, $\langle \hat m_{z}^{2}\rangle=4 \mu_B^2\langle \hat S_{z}^{2}\rangle$. The obtained temperature dependence of this property is shown in Fig.~\ref{fig:FeCo.DMFTmagMomSquared}.
One can see that the transition from ferromagnetic to PM phase can also be traced for $\langle\hat m_{z}^{2}\rangle$.
As it follows from Figure~\ref{fig:FeCo.DMFTmagMomSquared}, it has a nonlinear temperature dependence for T<2499 K and becomes constant at higher temperatures, where it reaches the values of 8.8 and 4.7 $\mu^2_B$ for Fe and Co, respectively.

We have also verified how the choice of $U$ and $J$ affects these results. 
It was found that changing $U$ values to $U_{\mathrm{Fe,Co}}$ = 3 eV leads to a slight increase of squared fluctuating magnetic moments on iron atom (to 7.25 $\mu_B^2$) and an unchanged value of 3.18 $\mu_B^2$ on cobalt atom. 
At the same time, decreasing the value of $J$ to 0.6 eV results in the decrease of these quantities on both atoms by 2.47 $\mu_b^2$ on Fe and 0.49 $\mu_b^2$ on Co. Such a sensitivity of the fluctuating moments to the choice of $U$ and $J$ is not unique to the present transition metal alloy \cite{anisimov2010electronic}.
The most important point is that the fluctuating moment on Co is substantial and this result is independent of the choice of Coulomb parameters.

Due to the symmetry breaking introduced by the density-density approximation, the fluctuations along various directions are different even above $T_c$ (see e.g. \cite{Belozerov2014}). 
As a result, the $\langle\hat m_{z}^{2}\rangle$ presented above can not be used to estimate the actual paramagnetic moments.
The latter can be extracted from paramagnetic susceptibility.

\begin{figure}[!h]
\includegraphics[width=\columnwidth]{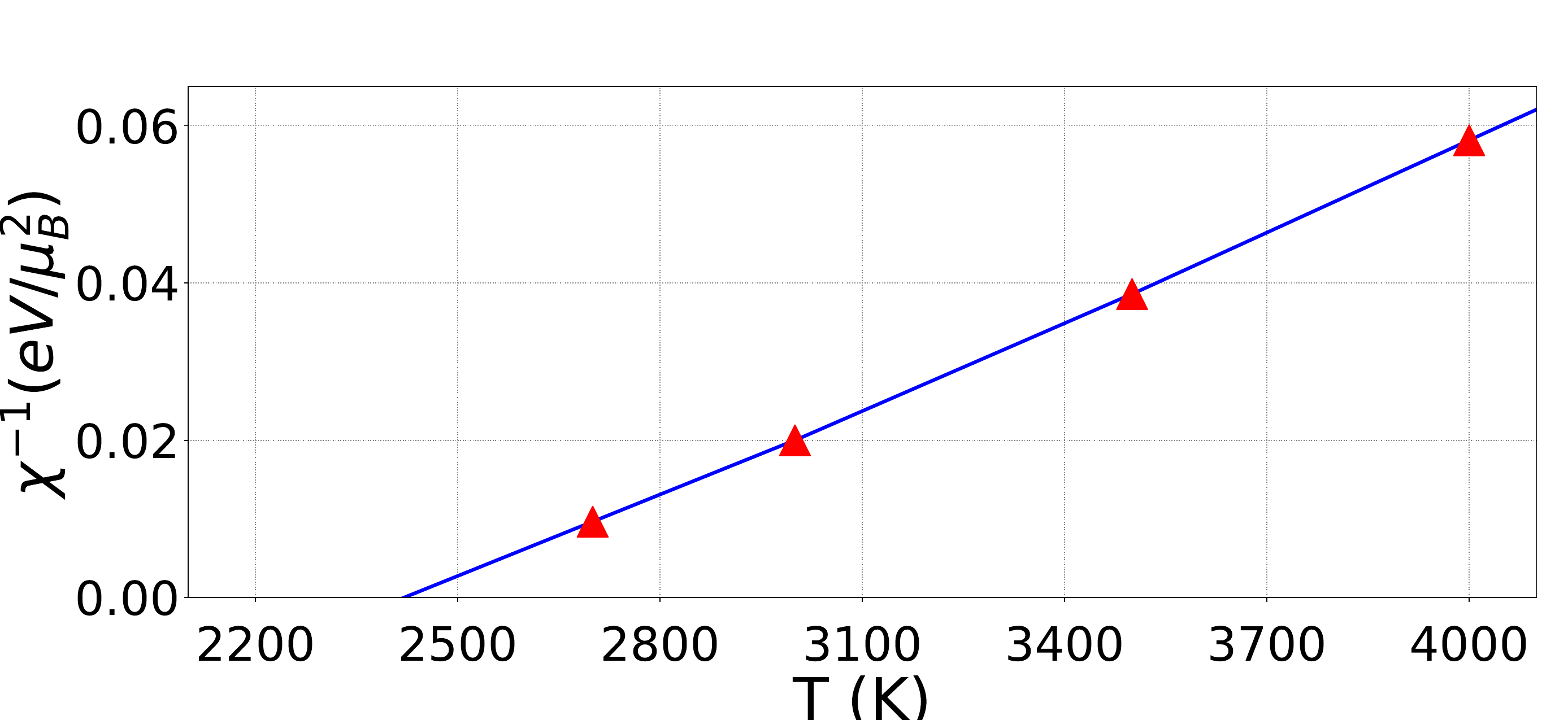}
\caption{Inverse uniform magnetic susceptibility in B2 FeCo alloy as a function of temperature from LDA+DMFT calculation.}
\label{fig:FeCo.DMFTchi}
\end{figure}

The uniform magnetic susceptibility was calculated directly in LDA+DMFT by computing the derivative of the magnetization, induced by an applied external field ($\chi(T)=\frac{\partial M}{\partial H}|_T$).
The obtained $\chi^{-1}(T)$ in FeCo is shown in Figure~\ref{fig:FeCo.DMFTchi}. 
For the temperatures above 2400 K it can be fitted according to the Curie-Weiss (C-W) law: $\chi^{-1}(T) = 3(T-\theta)/\mu^2_{eff}$, where $\theta$ is the Weiss constant and $\mu_{eff}$ is the effective paramagnetic moment. 
The extracted Weiss constant has a value of about 2454 K, which agrees well with the value of $T_c$ (2499 K) previously obtained from the calculated ordered moments as a function of temperature. 
The effective moment $\mu_{eff}$ from the C-W fit equals to $2.62 \mu_B$ with the contributions coming from Fe and Co being equal to $2.12 \mu_B$ and $1.69 \mu_B$, respectively. 
Thus, these results also confirm the presence of rather large Co moments (with similar values as in the ordered phase) existing above $T_c$.

The predicted values of the paramagnetic moments can be verified experimentally if the $T_c$ is reduced so that it is much lower than the chemical order-disorder critical temperature.
Based on first-principles calculations \cite{jakobsson13}, it has been suggested that the $T_c$ in the B2 FeCo alloy can be tuned by applying strain.
If this strategy is successful, then the total effective moment above $T_c$ can be potentially extracted from susceptibility measurements.
The element-resolved information can be obtained, for instance, from $K_{\beta}$ X-ray emission spectroscopy \cite{PhysRevB.13.929}.

\section{Conclusion}
By using an arsenal of first-principles methods and many-body approaches we explored the magnetic properties of FeCo alloy. Non-collinear LSDA calculations have revealed that $m_{\mathrm{Co}}$ is very sensitive to the magnetic configuration of the surrounding iron atoms.
This is related to an exceptionally strong inter-atomic exchange interactions between Fe and Co atoms as compared to intra-atomic exchange on Co atom.
The "bare" DLM calculations qualitatively support this picture and predict Co to be non-magnetic in the paramagnetic phase. Taking into account temperature-induced longitudinal fluctuations restores finite Co magnetization, but the values of $m_{\mathrm{Co}}$ are much smaller than in LSDA and show a very strong temperature-dependence.

The situation changes dramatically once Hubbard $U$ is explicitly added into the picture. Already DLM+$U$ calculations suggest a formation of substantial $m_{\mathrm{Co}}$ even in the completely uncorrelated disordered state.
More sophisticated LDA+DMFT calculations also predict the existence of a robust local magnetic moment on cobalt in the paramagnetic state and a reasonable estimate of the $T_c$.
Strong Coulomb correlations also result in a substantial mass enhancement and a pronounced non-Fermi-liquid behaviour of Fe $e_g$ states.

We argue that strong local correlations are necessary to properly describe finite-temperature electronic and magnetic properties of ordered FeCo alloy.
The LDA+DMFT approach thus provides the minimal physically sound model of this material.

\ack
A.G. acknowledges A. I. Poteryaev and  A. S. Belozerov for fruitful discussions, EUSpec COST Action and the Competitiveness Enhancement Program - CEP 3.1.1.2.$\Gamma$-17.
The work is supported by the Swedish Research Council (VR) and Swedish Foundation for International Cooperation in Research and Higher Education (STINT).
The computer simulations are performed on computational resources provided by NSC allocated by the Swedish National Infrastructure for Computing (SNIC).

\end{document}